\documentclass[%
aip,
rsi,
reprint,
superscriptaddress,
nofootinbib,
 amsmath,amssymb,
]{revtex4-2}
\usepackage{float}
\usepackage{graphicx}
\usepackage{ulem}
\usepackage[english]{babel}
\usepackage[utf8]{inputenc}
\usepackage{dcolumn}
\usepackage{bm}
\usepackage{xcolor}

\begin{document}

\preprint{AIP}

\title{Design and simulation of an Imaging Neutral Particle Analyzer for the ASDEX Upgrade tokamak}
\author{J. Rueda-Rueda}
\email{jrrueda@us.es}
\affiliation{University of Seville, Seville, Spain}
\author{M. García-Muñoz}%
\affiliation{University of Seville, Seville, Spain}
\affiliation{Centro Nacional de Aceleradores (CNA) CSIC, Seville, Spain}
\author{E. Viezzer}%
\affiliation{University of Seville, Seville, Spain}
\affiliation{Centro Nacional de Aceleradores (CNA) CSIC, Seville, Spain}
\author{P. A. Schneider}%
\affiliation{Max Planck Institute for Plasma Physics, Boltzmannstr. 2, 85748 Garching, Germany}
\author{J. García-Domínguez}%
\affiliation{Centro Nacional de Aceleradores (CNA) CSIC, Seville, Spain}
\author{J. Ayllon-Guerola}%
\affiliation{Centro Nacional de Aceleradores (CNA) CSIC, Seville, Spain}
\author{J. Galdón-Quiroga}%
\affiliation{Max Planck Institute for Plasma Physics, Boltzmannstr. 2, 85748 Garching, Germany}
\author{A. Herrmann}%
\affiliation{Max Planck Institute for Plasma Physics, Boltzmannstr. 2, 85748 Garching, Germany}
\author{X. Du}
\author{M. A. Van Zeeland}
\affiliation{General Atomics, San Diego, CA, United States of America}
\author{P. Oyola}
\affiliation{University of Seville, Seville, Spain}
\author{M. Rodriguez-Ramos}
\affiliation{Laboratory for Ion Beam Interactions, Ruđer Bošković Institute, Zagreb, Croatia}
\author{ASDEX Upgrade team}
\altaffiliation{See author list of H. Meyer et al. Nucl. Fusion \textbf{59} 112014 (2019)}
\noaffiliation

\date{11/01/2021}
\begin{abstract}
An Imaging Neutral Particle Analyser (INPA) diagnostic has been designed for the ASDEX Upgrade (AUG) tokamak. The AUG INPA diagnostic will measure fast neutrals escaping the plasma after CX reactions. The neutrals will be ionised by a 20 nm carbon foil and deflected towards a scintillator by the local magnetic field. The use of a neutral beam injector (NBI) as active source of neutrals will provide  radially resolved measurements while the use of a scintillator as active component will allow us to  cover the whole plasma along the NBI line with unprecedented phase-space resolution ($ < $ 12 keV and 8 cm) and fast temporal response (up to 1 kHz with the high resolution acquisition system and above 100 kHz with the low resolution one); making it suitable to study localised fast-ions redistributions in phase-space.

\end{abstract}

\maketitle

\section{\label{sec:Int}Introduction}

A detailed understanding of the fast-ion (FI) behaviour in the presence of magnetohydrodynamic (MHD) fluctuations is mandatory for achieving a good fast-ion confinement in future fusion devices. \cite{Fasoli2007} For this purpose, novel diagnostic techniques to measure the FI distribution in phase-space with Alfvénic temporal resolution are currently being developed. The Imaging Neutral Particle Analyser (INPA), already installed at the DIII-D tokamak \cite{Du2018,VanZeeland2019}, which is able to measure the radial position and energy of the confined FI population with fast temporal response response, is one example among these novel diagnostics.

This paper is structured as follows: the INPA working principle is explained in section 2. Section 3 presents the synthetic diagnostic while section 4 explains the influence of the different geometrical parameters on the detector performance. Section 5 presents the response of INPA to FI re-distributions due to magnetohydrodynamic (MHD) activity.

\section{\label{sec:work}Working principle}
\begin{figure}[h!] 
	\includegraphics[width=0.48\textwidth]{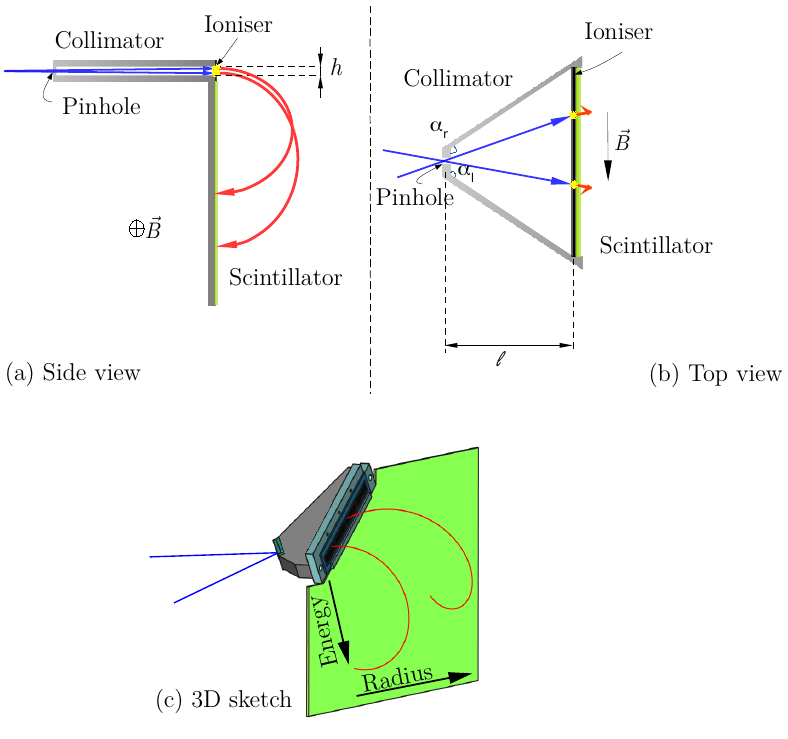}
	\caption{\label{fig:collimator} Scheme of the INPA working principle. In blue the trajectory of the neutral particle while in red, the one of the ion. Main dimensions of the collimator are also highlighted.}
\end{figure}

\begin{figure}[t]
	\includegraphics[]{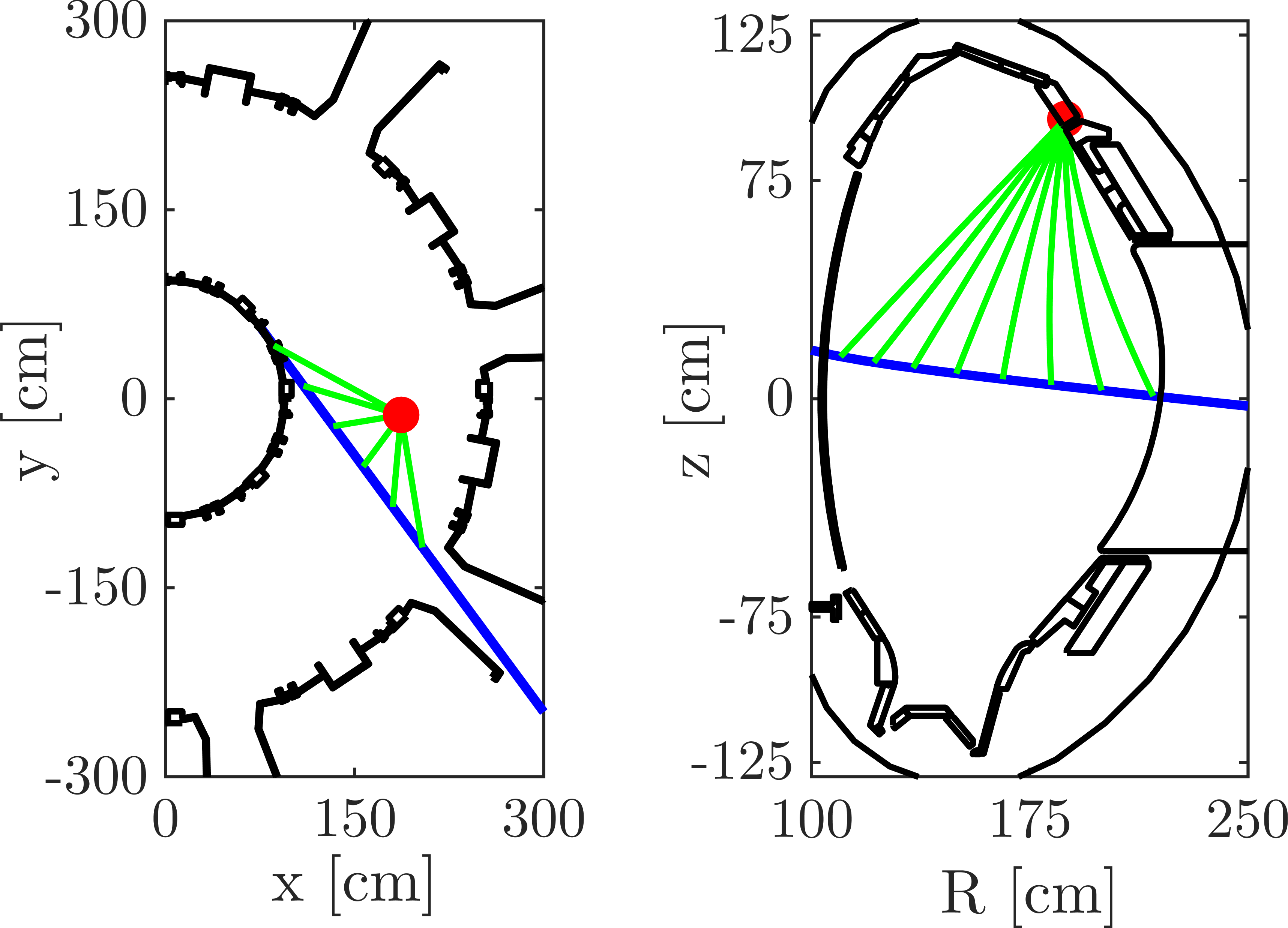}
	\caption{\label{fig:INPA_LOS}Relation between line-of-sight (LOS) and radial birth position. As neutral particles travel along a straight line, and assuming they all are born along the NBI line, there is a one-to-one relation between the LOS (velocity direction of the neutral) and its radial birth position. In blue the NBI line, in green some of the explored LOSs and in red the location of the INPA head.}
\end{figure}

INPA combines the already working principles of neutral particle analysers (NPA) \cite{NPA} and fast-ion loss detectors (FILD) \cite{Garcia-Munoz2009} to provide the energy and location of the confined fast-ion population with fast temporal response. Similar to NPA systems, INPA analyses neutral particles produced in charge exchange (CX) reactions. These particles are not confined by the magnetic field and retain all the information of the movement of fast-ions, as no significant exchange of energy or momentum occurs during the CX reaction \cite{Hutchinson_book}. After being collimated, neutrals are ionized by an ultra-thin (20 nm, for the AUG set-up) carbon foil and deflected towards a scintillator plate, as can be seen in figure \ref{fig:collimator}. The strike position of the particles in the scintillator will be given by their energy and pitch ($ \lambda = -v_\parallel/v $, where $ v_\parallel$ is the velocity along the magnetic field) and the local magnetic field of the tokamak. The use of a neutral beam injector (NBI) as active source of neutrals gives the possibility of relating, via modelling, the measured pitch  of the neutrals with their radial birth position; and hence, with the location of the confined FI. This is possible thanks to the geometric relation between the velocity orientation of the measured neutral and its radial birth position.Figure \ref{fig:INPA_LOS} shows an example of this relation for the case of AUG. The in-vessel positioning of the diagnostic opens up a broader flexibility in selecting the lines-of-sight (LOS) compared to conventional electromagnetic NPA installed out-vessel and using their own electromagnetic fields\cite{NPA}. The use of a scintillator will enable the simultaneous exploration of a wide region of the plasma, which would not be covered by individual LOS. Light emitted by the scintillator is collected by a series of lenses and guided to a Phantom camera and a set of photomultipliers (PMT). The camera allows us to obtain high resolution images of the scintillator while the PMTs sacrifice this resolution to gain faster temporal response.

\section{\label{sec:INPASIM} Synthetic diagnostic}
A synthetic INPA diagnostic has been developed to study the feasibility for the installation of an INPA at AUG and to optimize its design. The synthetic INPA is based on FIDASIM \cite{FIDASIM_WH,FIDASIM_ULTIMATE} output and the INPASIM code, which has been developed during this work. Given the magnetic equilibrium, plasma profiles, fast-ion distribution function and the detector geometry, FIDASIM calculates the flux of neutral particles, coming from CX reactions, with a Monte Carlo (MC) approach\footnote{FIDASIM (Fast-ion D$\alpha$ Simulator) main purpose is to simulate D$\alpha$ radiation, but can also track CX neutrals up to an NPA.}. To this end, a collisional-radiative model is solved\cite{FIDASIM_WH}. FIDASIM has been extensively verified against experimental data at the DIII-D\cite{Luo2007}, AUG\cite{FIDASIM_PS} and TCV\cite{Geiger2017} tokamaks, the LHD\cite{Fujiwara2020} stellarator and other devices\cite{FIDASIM_ULTIMATE}. Using the FIDASIM output (velocity-space resolved neutral densities) as input, the INPASIM code calculates the synthetic signal, the resolution, as well as the instrument function to perform tomographic reconstructions as for FILD \cite{Galdon-Quiroga2018_tomo}.
 \begin{figure}[b!]
	\includegraphics{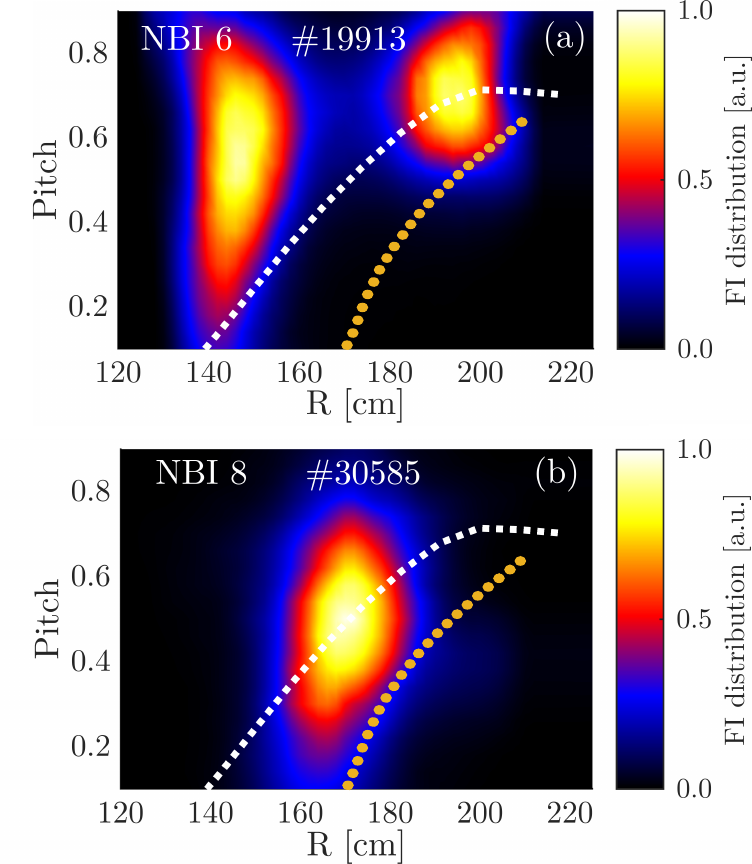}
	\caption{\label{fig:fbm_6_8} Pitch profile explored by the INPA diagnostic. (a) Projection in the $(R,\lambda)$-space of the fast-ion distribution, calculated with the TRANSP \cite{TRANSP_ORIGINAL,TRANSP_CODE} code, using NBI\#6 and shot \#19913. The dotted (squares) white line corresponds to a geometric calculation of the explored pitch profile based on the NBI geometry and the magnetic equilibrium, the orange line (circles) corresponds to the trapped-passing boundary at the height of the magnetic axis. (b) Equivalent for NBI\#8 and shot \#30585. The acceptance of INPA is around $ \pm 0.1 $ on the pitch profile shown.}
\end{figure}
INPASIM is divided in two independent sections: calculation of the signal / instrument function and determination of the scintillator strike map. In the former, the code tracks the FIDASIM markers inside the diagnostic head until they collide with the collimator or scintillator, both considered as 3D elements. In the latter, MC markers with given energy and pitch are launched at the pinhole in order to create a map which relates the ($ R,E $)-space with the strike position of the particle in the scintillator. For both sections, scattering and energy-loss in the carbon foil are modelled following SRIM \cite{SRIM} simulations, the ionization yield in the carbon foil follows DIII-D INPA modelling\cite{Du2020} and the scintillator yield follows Birk's model as applied in absolute measurements of fast-ion losses at AUG\cite{MauriPhD,Joaquin_absolute_fildsim}. This model predicts the number of photons emitted by the scintillator per incident ion. After the scintillator emission is calculated, INPASIM forms the camera image considering the transmission factor of the optical system and introducing a 2D Gaussian function to mimic the finite focusing of the optics; both based on Zemax simulations.

\section{\label{sec:optimization}Design of the INPA diagnostic}

 \subsection{\label{position_head}Selection of the position to place INPA at AUG}

The selection of the position for the INPA diagnostic is a compromise between four factors: phase space coverage, signal level (attenuation), available space inside AUG and resolution. The orientation of the diagnostic will determine which velocity directions can be measured, therefore, which region of the FI phase-space can be probed. Special care has been placed in matching the pitch profile of the slowing-down fast-ion distribution created by NBI\#8 and NBI\#6 of AUG; which are typical examples of the on- and off-axis FI distribution achievable in AUG via NBI heating. A comparison of the pitch profiles of these distributions and the one INPA will explore can be found in figure \ref{fig:fbm_6_8}. Here, the pitch profile explored by the INPA diagnostic is highlighted with the white dashed line and the trapped-passing boundary with an orange one. Notice that INPA will be mainly sensitive to passing ions.

\begin{figure}
	\includegraphics{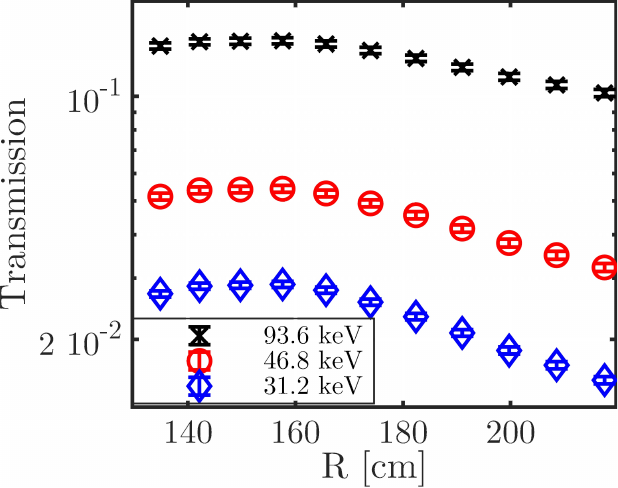}
	\caption{\label{fig:transmission} Fraction of CX neutrals created in reactions with the active source which reach the detector pinhole, as a function of their radial birth position. It corresponds to a density of $n_e (0) = 6.5 \cdot 10^{19} \mathrm{ m^{-3}}$. Calculated with FIDASIM.}
\end{figure}

The pinhole position to achieve this good overlapping is found in sector 16 of AUG, the sector closest to NBI\#3 which provides the active source of neutrals. Therefore the path that CX neutrals should travel inside the plasma and the corresponding losses due to re-ionization are minimised. The transmission factor for a \textit{high} density case ($n_e (0) = 6.5 \cdot 10^{19} \mathrm{ m^{-3}}$) can be found in figure \ref{fig:transmission}.


\subsection{\label{sec:geom_head}Collimator and energy resolution}
\begin{figure}[b]
	\centering
	\includegraphics{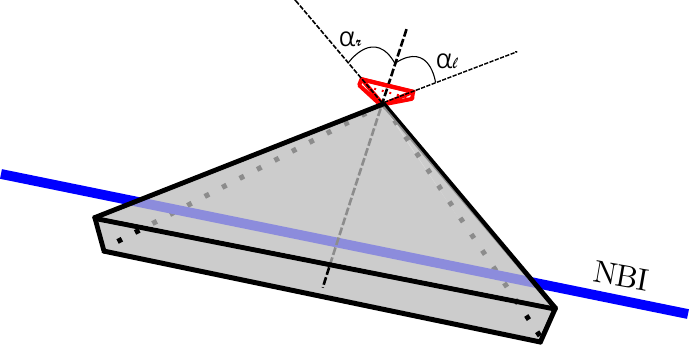}
	\caption{Effect of the collimator angles in the INPA view. In red a 3D sketch of the INPA collimator, in shaded gray the solid angle covered by the INPA while in blue the NBI line. The bigger are the angles $\alpha_r$ and $\alpha_r$ the larger is the solid angle covered by the diagnostic.}
	\label{fig:figureb}
\end{figure}

\begin{figure} 
	\includegraphics{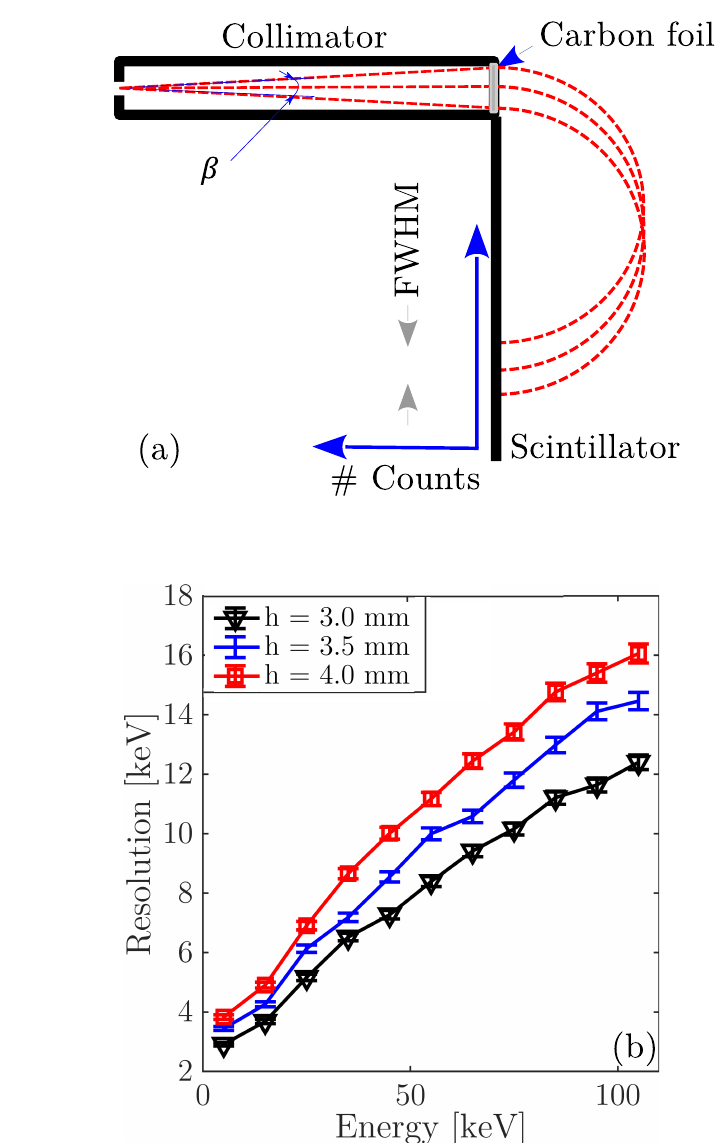}
	\caption{\label{fig:energy_res}Energy resolution. (a) influence of the collimator height on the energy resolution; several possible trajectories of particles with the same energy are plotted. The larger the height of the carbon foil, the wider the distribution of impacts on the scintillator. (b) Energy resolution (FWHM) of INPA for different values of the collimator height. Here, the magnetic field is 2.5 T on the axis.}
\end{figure}

The basic shape of the INPA collimator is sketched in figure \ref{fig:collimator}. The collimator length, $l$, is fixed by the available distance to the first wall but all the other parameters are free to be modified. The aperture angles of the collimator, $\alpha_l$ and $\alpha_r$, control its acceptance in the direction of the NBI, as can be seen in figure \ref{fig:figureb}. They are  $30^o$ and $40^o$, respectively. These values enable a coverage of the region from R = 1.35 m up to the outer separatrix, approximately at R = 2.16 m. The acceptance of the diagnostic in the perpendicular direction, denoted by $\beta$ in figure \ref{fig:energy_res}(a), dominates the energy resolution. This acceptance is controlled by the pinhole size, collimator length and its height, $h$. Decreasing the pinhole radius or $h$ will improve the energy resolution but will reduce the signal level. As decreasing the pinhole radius reduces the neutral influx quadratically, $h$ is the chosen factor to pursuit the desired value of energy resolution, as it only affects the signal level linearly. A detailed comparison of the energy resolution quantities for different values of $h$ can be found in figure \ref{fig:energy_res}(b). Values below $h=3$ mm are not considered in order to maintain a flux high enough to reach Alvénic time scale with the PMTs and values above $h=4$ mm are avoided to keep a good energy resolution. The value $h=3$ mm was selected as final choice, looking for the best energy resolution. Possible scattering caused by the carbon foil will also deteriorate the resolution, but the small thickness of the carbon foil (20 nm) makes this scattering unimportant for NBI injection energies at AUG.


\subsection{\label{sec:Radial_res}Radial resolution}

\begin{figure}[h] 
	\includegraphics[width=0.3\textwidth]{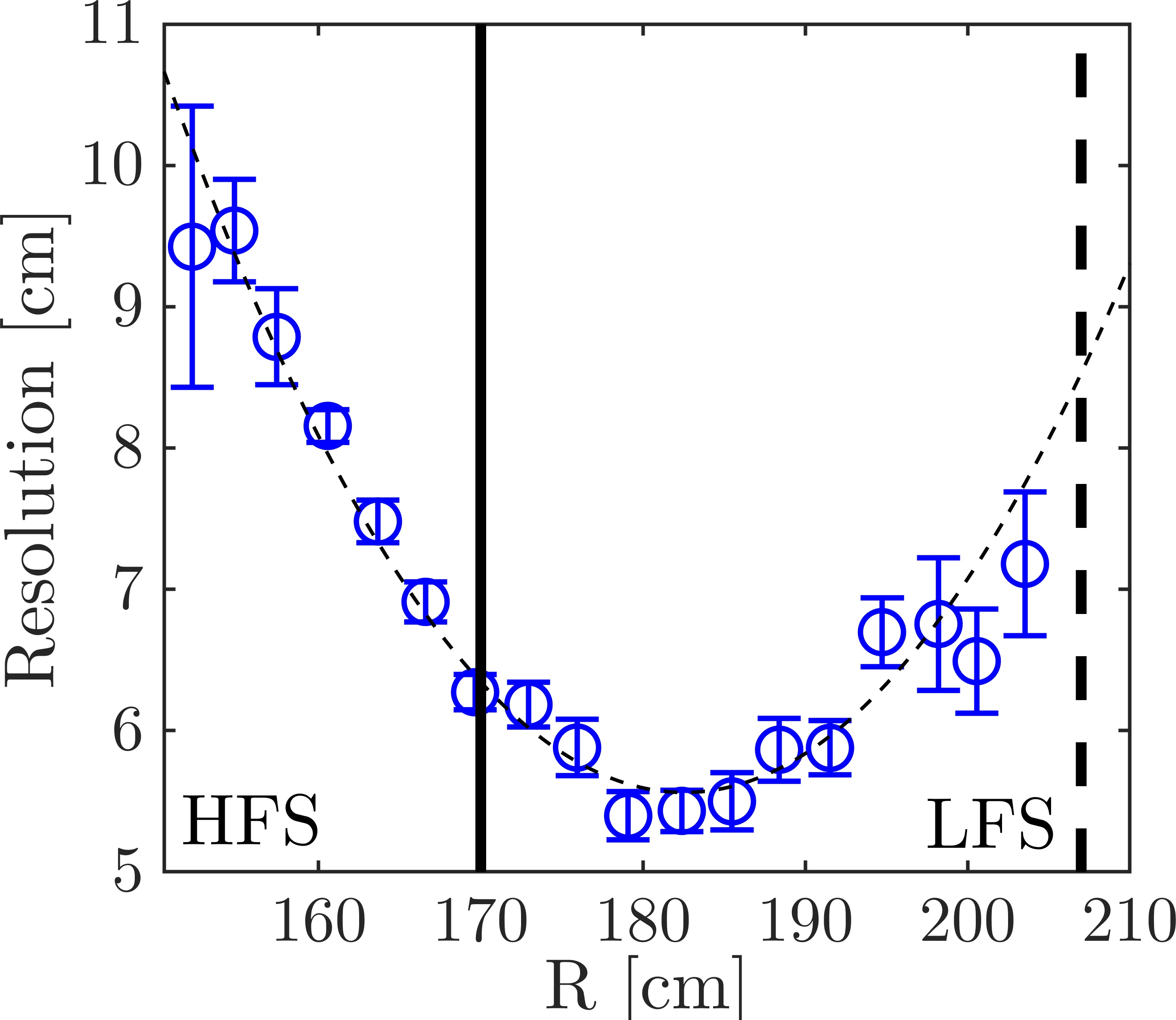}
	\caption{\label{fig:radial_res} Radial resolution of INPA. Calculated for shot \#30585.The magnetic axis position is indicated with a solid black line while the outer separatrix position with a dotted one.The high field side (HFS) and low field side (LFS= are also indicated.}
\end{figure}

Two factors dominate the INPA radial resolution:  the NBI source diameter which acts as active neutral source and the diameter of the pinhole. Synthetic signals calculated by the INPASIM code have been used to estimate the radial resolution. To this end, the actual birth position of the markers has been compared with the position given by the strike map. The full width half maximum (FWHM) of the structures resulting of this comparison, can be found in figure \ref{fig:radial_res}, where a parabolic fit has only been included as a guideline to the eye. A priori, there is no model which justifies that the radial resolution must have a parabolic dependence with the major radius.

\subsection{\label{sec:temp}Temporal resolution}

\begin{figure}[b] 
		\includegraphics[width=0.3\textwidth]{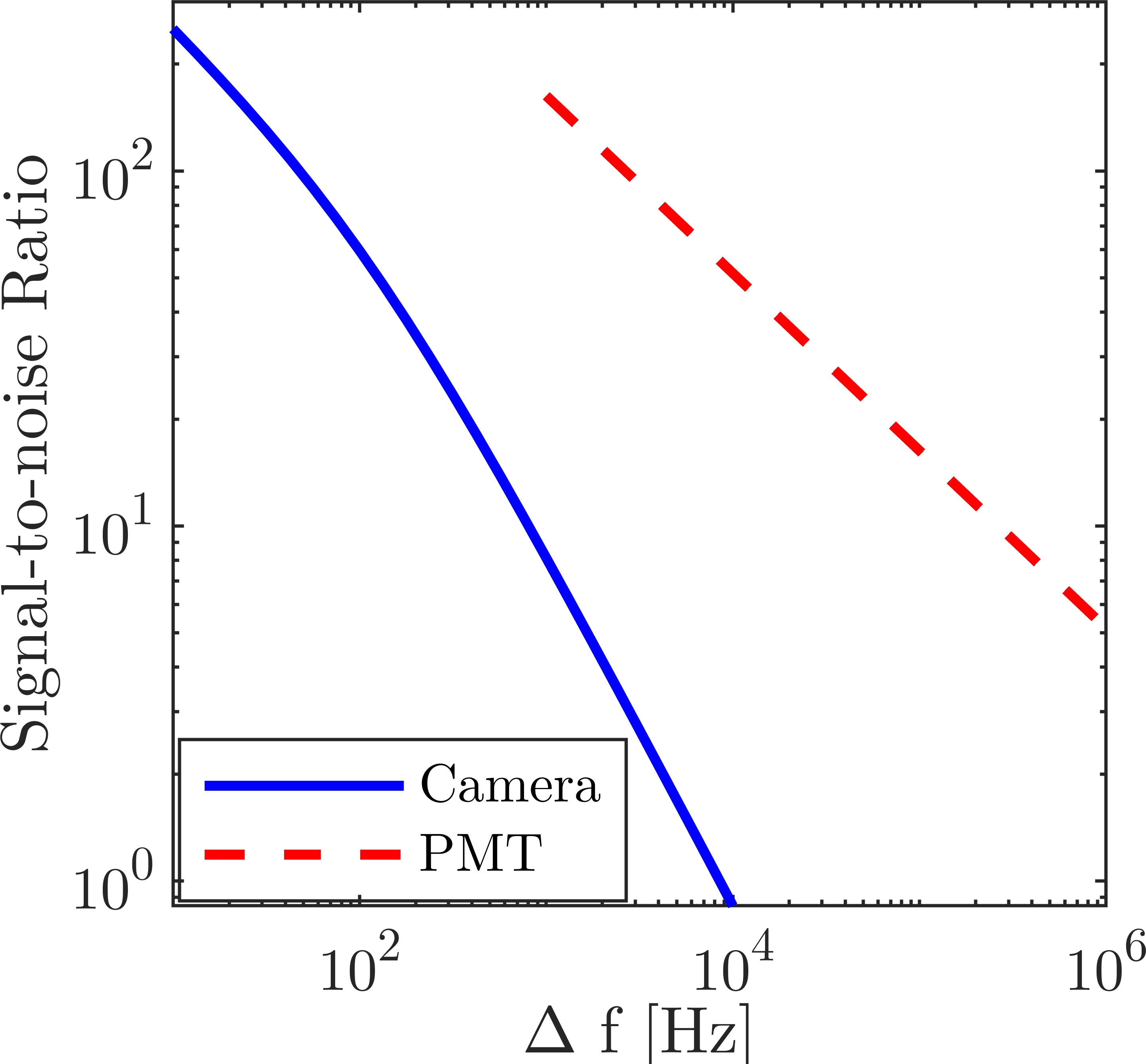}
		\caption{\label{fig:SNR_}Signal-to-noise ratio of the acquisition systems. The calculation has been performed for a case with a core plasma density of $n_e(0) = 6.5 \cdot 10^{19} \mathrm{m^{-3}}$. Calculated with INPASIM.}
	\end{figure}

	In general, the signal-to-noise ratio (SNR) at a photo sensor can be written as \cite{Hamamatsu-Photonics2007}:
	\begin{equation}
		\begin{aligned}
			\mathrm{SNR} ={} & \frac{n_c}{\sqrt{\sigma(n_c)^2+\sigma_{reading}^2} } = \\
			= &\frac{QE\phi}{\sqrt{2QE\phi\Delta f+(2\Delta f)^2\sigma_{reading}^2 }}
		\end{aligned}
	\end{equation}
	where $n_c$ represents the number of electrons created in the photo-cathode of the sensor (camera sensor or PMT) assumed to follow a Poison distribution, $\sigma$ the noise, $QE$ the quantum efficiency, $\phi$ the photon flux and $\Delta f$ the measurable bandwidth. Notice that the dark current term was neglected. This expression is plotted in figure \ref{fig:SNR_} for the fluxed predicted by INPASIM in a case of a core plasma density of $n_e(0) = 6.5 \cdot 10^{19} \mathrm{m^{-3}}$. Taking a SNR of 10 as an adequate value, up to 1 kHz of bandwidth could be achieved using the Phantom camera (which allows to have the energy and radial resolutions presented above) and above 100 kHz could be achieved with the PMT array.
\begin{figure*}
	\centering
	\includegraphics[width=\linewidth]{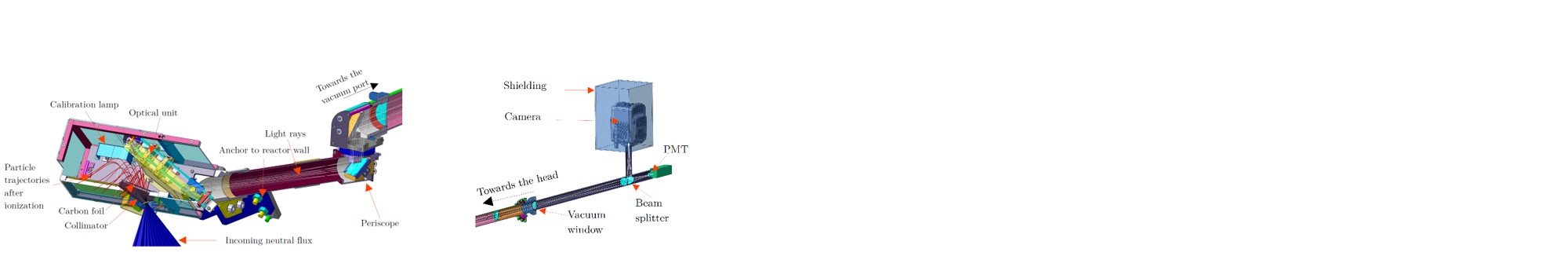}
	\caption{Section of the CAD design of the AUG INPA diagnostic. (a) In-vessel components, (b) out-vessel components.}
	\label{fig:cad}
\end{figure*}

	\subsection{\label{mech_design}Mechanical design}
	An overview of the mechanical design of the AUG INPA diagnostic is shown in figure \ref{fig:cad}. As can be seen, inside the detector head not only the first optical elements and the scintillator are located but also a calibration lamp, which allows to check the alignment of the optical components between shot days. The head is connected with the port window with a periscope. This periscope allows not only to meet the spatial boundary conditions but also provides the system with the necessary degrees of freedom to accommodate small deviations between CAD and reality during installation. In the out-vessel region, a beam splitter divide the photon flux and redirect it towards the Phantom camera and the PMT array.
 

\section{\label{sec:fluc}Response of INPA to MHD fluctuations}
\begin{figure*}
	\centering
	\includegraphics{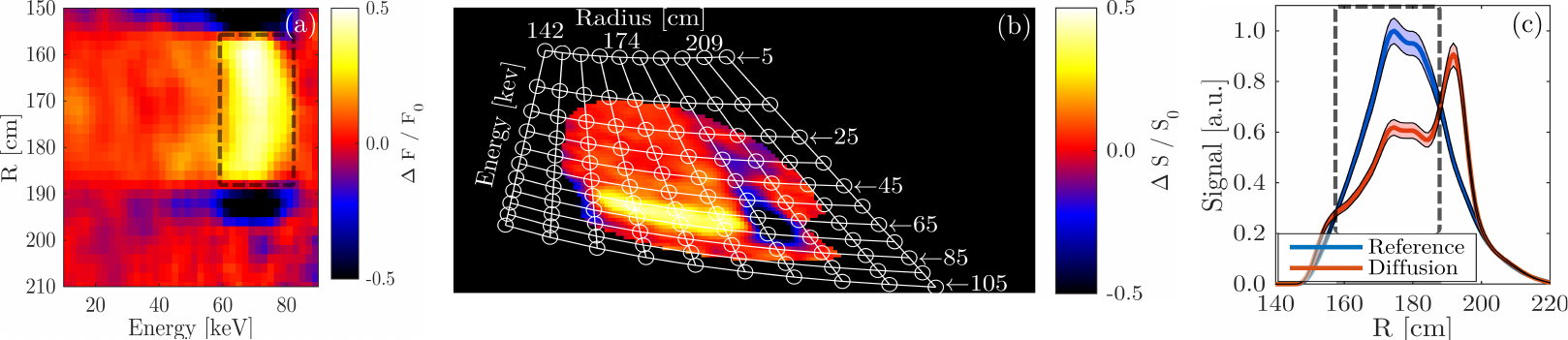}
	\caption{(a) Differences in the FI distribution due to the assumed anomalous diffusion. Only the mid-plane region (z = 0) has been plotted. Calculated with TRANSP. (b) Relative changes in the scintillator yield. To avoid simulation noise, only pixels which have more than 5\% signal have been plotted. (b) Cut at 72 keV. Shaded areas represent a standard deviation of 5\%, expected upper limit of measurement noise. The grey dashed frames indicate where the anomalous diffusion was applied.}
	\label{fig:redistribution}
\end{figure*}

	A complete review of the INPA response to different plasma scenarios and FI redistributions is out of the scope of this article and will be presented in a follow-up paper. Here, we present only one example to show the capabilities of the diagnostic. In this case, we simulate the response of INPA to a localised redistribution. An ad-hoc anomalous diffusion coefficient has been inserted in TRANSP to mimic the effect of a localised MHD fluctuation in the FI distribution. A maximum value of the anomalous diffusion coefficient of 0.5 $ \mathrm{m ^2/s} $\cite{VanZeeland2011,Heidbrink2008} has been set; and with an extension of 0.3 in the $ \rho_t $ space ($\rho_t$ is the normalised toroidal flux radius), which is in concordance with the size of the different poloidal modes of a TAE measured and simulated at DIII-D \cite{White2010,VanZeeland2006}. 250 ms were simulated to give enough time for the FI to slow down. As mentioned in \cite{Heidbrink2017}, this simulation scheme cannot reproduce the precise interaction between the MHD instabilities and fast-ions but gives an estimate of the particle transport. The anomalous diffusion has been selected to be constant in the interval $ \rho_t \in(0,0.3) $ and zero otherwise. In energy, the coefficient takes a Gaussian shape centred at 72 keV, with a FWHM of 10 keV, to mimic a narrow resonance. Anomalous diffusion has only been applied to passing particles.
	The relative difference in the distribution function can be seen in figure \ref{fig:redistribution}(a). Notice that in the region between R = 1.55 to 1.85 m, the FI density is smaller when the diffusion is activated, as could be expected (the range where the diffusion was applied is highlighted in grey). The differences in the scintillator signal can be seen in figure \ref{fig:redistribution}(b). Note how it agrees with the differences in the FI distribution, inside the resolution of the diagnostic. In figure  \ref{fig:redistribution}(c) the difference along the constant line of 72 keV can be seen. Both curves will be easily distinguishable, even with the assumed 5\% of noise. During this comparison, the plasma profiles have been considered to remain constant. 

The INPA diagnostic is, in its current configuration, not expected to be sensible to a redistribution of trapped particles; because the explored pitch profile is basically always in the passing region of the phase space (see figure \ref{fig:fbm_6_8}). Only for regions close to the separatrix, the trapped-passing boundary enters in the region explored by INPA and the diagnostic becomes sensible to trapped particles. Therefore, in general, only passing ions or trapped ions which become passing and enter the INPA field of view after the diffusion will be measurable.

\section{Conclusions}
An imaging neutral particle analyser has been designed and optimised for the ASDEX Upgrade tokamak. This diagnostic will allow measurements of the distribution of supra-thermal particles in energy and radius with  good resolution and fast temporal response simultaneously, complementing the AUG suite of fast-ion diagnostics to obtain a complete understanding of the dynamics and transport of supra-thermal particles.

The final design of the diagnostic features an energy resolution of 12 keV for 100 keV ions and a radial resolution below 8 cm at the low field side of AUG; with a temporal response of 1 kHz. If the fast acquisition system (with low spatial resolution) is used, the response is increased to above 100 kHz.

\section{Acknowledgements}
This project has received funding from the European Research Council (ERC) under the European Union’s Horizon 2020 research and innovation programme (grant agreement No. 805162) and the Spanish \textit{Ministerio de Ciencia, Innovación y Universidades} (grant FPU19/02486).

The data that support the findings of this study are available from the corresponding author upon reasonable request.

\bibliography{apssamp}

\end{document}